# Sensor properties of Pd doped $SnO_2$ nanofiber enshrouded with functionalized MWCNT

Cihat Tasaltin[*]

*TÜBİTAK Marmara Research Center, Materials Institute, P.O Box 21*

*41470 Gebze, Kocaeli, Turkey*

**[*] Corresponding Author**

Dr. Cihat Tasaltin

Tel: +90 262 677 3042

Fax: +90 262 677 2309

E-mail address: cihat.tasaltin@tubitak.gov.tr

**Abstract** In this work, tin oxide (SnO2)/multi-walled carbon nanotube (MWCNT) composite based ethanol sensor that exhibits fast response/recovery behaviour, large sensitivity, and good selectivity was demonstrated. First, Pd doped SnO2 (SnO2-Pd) nanofibers was prepared on the interdigitated electrodes (IDE) using electrospinning technique, followed by calcination at 600 $^0$C for 2 h. Then, carboxylic acid modified multi-walled carbon nanotube (MWCNT-COOH) was dropped on the nanofibers and allowed to dry at RT. Next, the thin films were characterized by field emission scanning electron microscopy (FE-SEM), transmission electron microscopy (TEM), X-ray diffraction (XRD) and thermogravimetric analysis (TGA). Finally, chemical sensing behaviors of the sensors were analyzed against to ethanol (ETH), toluene (TOL), trichloroethylene (TCE), acetone (ACE) and binary mixing with water. It was demonstrated that increasing the temperature and gas concentration has led to increased sensor response for SnO2-Pd nanofiber system. However, there was no general rule found for MWCNT-COOH coated SnO2-Pd nanofiber system. A novel sensing mechanism was offered for the composite structure.

# 1. Introduction

Oxide semiconductor gas sensors undergo resistance change upon exposure to reducing gases by oxidative interactions with the negatively charged chemisorbed oxygen. The gas sensing characteristics such as gas response, responding speed, and selectivity are greatly influenced by the surface area, donor density, agglomeration, porosity, acid–base property of the sensing material, the presence of catalyst and the sensing temperature[1]. Tin oxide ($SnO_2$), n-type wide-gap semiconductor, is regarded as an important functional material for gas sensor applications owing to its small size, lightweight and simple fabrication procedures[2]. In addition, $SnO_2$ are frequently doped with noble metals such as palladium (Pd), platinum (Pt) and gold (Au) to promote the sensing properties such as selectivity, sensitivity and lower working temperature[3]. The form of sensing material is also another parameter for the sensors. For instance, one dimensional (1-D) structures such as nanofiber, nanobelt and nanowire has many advantages during sensing since 1-D structure can provide excellent electron transfer properties due to their large surface area to volume ratio, high porosity and higher gas diffusion area [4]. Among the 1-D materials, nanofibers have attracted great attention due to the simplicity in preparation.

Electrospinning is a widely utilized simple and economical method for producing nanofibers. In the past years, it was mainly used for preparation of pure organic polymer nanofibers[5, 6]. Recently, functional ceramic nanocomposite fibers, such as CuO, $SnO_2$, $TiO_2$, ZnO, $CeO_2$, $Ta_2O_5$, $Nb_2O_5$ and $TaNbO_5$ were prepared via electrospinning of metal precursor incorporated polymer solution [6-9]. In a typical electrospinning process, nanofibers are produced by applying high voltage to the viscous polymer solution [7]. Under the effect of a strong electric field, a droplet is formed at the end of metal tip. The droplet experiences two major types of electrostatic force: a Columbic force exerted by the external electric field and an electrostatic repulsion between surface charges [10]. With these electrostatic interactions, the droplet is ejected into a conical object (known as Taylor cone). When the applied voltage reaches a critical value, the electrostatic forces overcome the surface tension of the solution and an electrified jet is produced, which in turn led to formation of continuous nanofibers [7]. Nanofibers, which were produced using electrospinning technique, have been used for various applications such as gas sensor, filtration, biomaterial devices, tissue engineering, and photovoltaics [11-13].

Up to date, doped and undoped $SnO_2$ in various forms such as nanoparticle, nanofiber and nanohollow have been synthesized and used for detection of ethanol, $H_2$ and $NH_3$ by various research groups. In addition, Pd doped $SnO_2$ has been used for CO sensors by Gaidi *et al*[14]. Regardless of doping, $SnO_2$ is known to be very sensitive to moisture. Interaction of $SnO_2$ with the humidity was explained in detail in previous works by Heiland and Barsan [15-18]. According to the reports, there are two main significant mechanisms that play key role between the humidity and $SnO_2$. Firstly, water molecule can be adsorbed by the $SnO_2$ surface either as a molecule or as a hydroxyl group. If adsorption happens below 400 $^0$C, no effect on conductivity is observed. If adsorption occurs via chemisorption of hydroxyl groups, the resistance is reduced and this species is not completely desorbed even at 600 $^0$C. Heiland suggests two interaction mechanisms of water molecule; 1) dissociation or 2) reduction. In the former one, water molecules adsorbed by the surface results in hydrogen bonding and electron release. In the latter one, water molecule with two hydroxyl groups, rooted one including lattice oxygen another one bound to lattice tin. These phenomena can be expressed by the following equations[15].

$$H_2O + Sn_{lat} + O_{\ lat} \leftrightarrow (H_2O - Sn_{lat}) + O_{lat} - H + e \qquad \text{(eq. 1)}$$

$$H_2O + 2Sn_{lat} + O_{\ lat} \leftrightarrow 2(H_2O - Sn_{lat}) + V_0 \qquad \text{(eq. 2)}$$

On the other hand, small molecules such as methane, acetic acid and ethanol have an interaction with $SnO_2$ based on decomposition and oxidation processes [15]. Therefore, to obtain distinct sensor responses against to VOCs, $SnO_2$ has been doped by various metals. Conductivity of these materials has been discussed in the context of an activated tunneling model as defined in Eq. 3.

$$\sigma \approx \exp(-\beta\delta)\,exp(\frac{-E_a}{KT}) \qquad \text{(eq. 3)}$$

where $\beta$, $\delta$, $E_a$ and $k_b$ represents the tunneling decay constant, edge-to-edge separation of the metal cores, activation energy, Boltzmann constant, respectively. The first term defines the tunneling current between the neighboring of sensing material, which decreases with increasing particle distance [19]. The second term describes the thermal activation of charge carriers in which Arrhenius term is inversely proportional to the permittivity of the organic matrix. According to the equation, conductivity is also dependent of the dielectric constant of the analyte which was ascribed to a change in permittivity of the matrix as the dominating component of sensing mechanism [20, 21].

In this work, it was attempted to prepare Pd doped $SnO_2$ ($SnO_2$-Pd) nanofiber enshrouded with MWCNT-COOH layers to block the interference of humidity and other gases on sensor responses. To fabricate $SnO_2$-Pd nanofiber, electrospinning method was utilized. It is well known that, resistivity of MWCNT-COOH layer increases with the humidity because the adsorbed water molecules inside the MWCNT-COOH layer makes it difficult for the electrons to transport between the nanotubes [22, 23]. Other advantage of the MWCNT-COOH layer could be the effect of defects on the nanotubes in the sensing of polar molecules such as water and ethanol. Watt reported oxygen containing defects withdraws electron density from the nanotubes which increases the number of hole carriers and makes the nanotubes p-type with lower resistivity. Upon the adsorption of water molecules on the oxygen containing defects, electron-withdrawing power of the electronegative functional groups (i.e. carboxyl and epoxide) reduces, which in turn led to increased resistance [24].

## 2. Materials and methods

### 2.1 Materials and sensor fabrication

200 mg $SnCl_2·2H_2O$ (98 %, Sigma-Aldrich) was dissolved in 5 ml methanol and stirred for 1 h, followed by adding 200 mg Polyvinylpyrrolidone (PVP, Mw: 1.300.000 g/mol, Sigma-Aldrich) to the solution and stirring for 1 h (weight ratio 1:1). For the preparation of Pd doped $SnO_2$ nanofiber, 8 mg $PdCl_2$ (99,99 %, Sigma-Aldrich) was added to the above solution and stirred for 1 h. Subsequently, the resulting solution was loaded into a plastic syringe and electrified using a high-voltage DC supply. As demonstrated in Fig. 1, the electrospinning equipment is a two-compartment setup with a sample holder that rotates at a velocity of 1.000 rpm, exposing the IDE to the positive electrospray mist and a negative discharge cloud. The coating voltages were adjusted to ~+3,5 kV for the needle and ~-1,5 kV for the tungsten tip [25, 26]. After the electrospinning process, the nanofibers were subjected to calcination at 600 $^0$C in air for 2 h to remove the PVP and crystallize the $SnO_2$. To measure the sensor properties, an interdigitated electrode (IDEs) which consists of titanium finger pairs on $Si/SiO_2$ substrate with the following dimensions; 10 μm electrode width, 10 μm spacing and 2 μm electrode thickness were used.

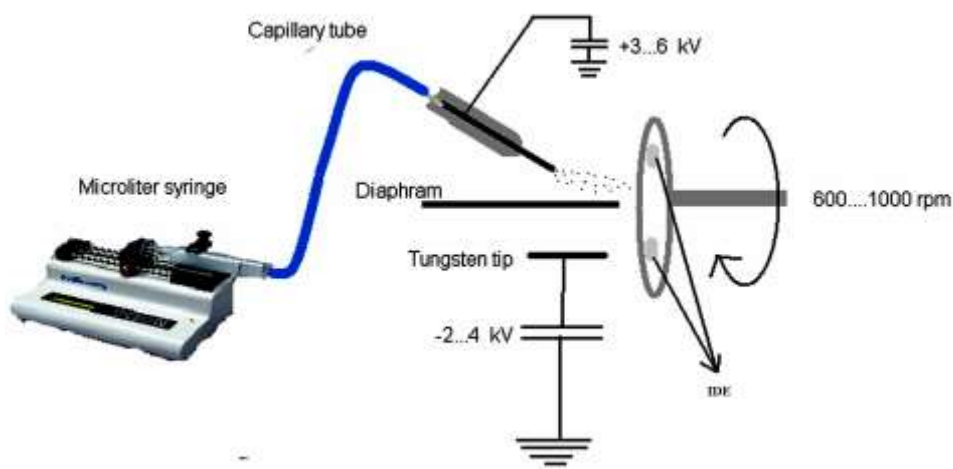

**Figure 1.** The electrospinning system[26]

Finally, carboxylic acid modified multi-walled carbon nanotubes (MWCNT-COOH, Nanoshel, US) solution was prepared in methanol. 10-15 µl portions was simply dropped on SnO$_2$ layer and allowed to dry at RT.

**2.2 Characterization**

Synthesized SnO$_2$-Pd nanofibers were characterized by X-ray diffractometer (XRD, Shimadzu XRD-6000), transmission electron microscopy (TEM, JEOL 2100) and thermogravimetric analysis (TGA, Seiko Exstar 6300) while the morphology of the sensing layers was analyzed by field emission scanning electron microscopy (FE-SEM, JEOL 63335F)

**2.3 Sensor measurement**

Chemical sensing behaviors were analyzed against polar (water, propanol and ethanol) and nonpolar (hexane, toluene, trichloroethylene and chloroform) volatile organic compounds (VOCs). The gas stream containing VOC vapor was generated from cooled bubblers that were immersed in a thermally controlled bath with synthetic air as the carrier gas. The gas stream saturated with the analyte was diluted with pure synthetic air to adjust the gas concentration to the desired amount by using computer driven mass flow controllers (MKS Instruments Inc., USA) at a constant flow rate of 200 ml/min. Typical experiments consisted of repeated exposure to the analyte gas (10 min) and a subsequent purging with pure air (10 min) to reset the baseline. The measurements were performed under individual and binary mixtures of VOC and humidity while the component of gas concentrations were varied in the range of 100–5000 ppm (Table 1). The binary measurements were taken under constant humidity and changing VOC concentrations. Meanwhile, the humidity levels were kept at 0, 30, 50 and 75 % relative humidity (RH) for each binary measurements. The temperature of the sensor chip was kept at 200, 250 and 300 $^0$C with the help of temperature controller (Lake Shore, USA). Electrical resistances (DC) were measured with a programmable electrometer (Keithley 617). Instruments were controlled and read by computer using a GPIB interface.

For a better comparison, the responses of different vapor pressures' relative concentrations $p_i/p_{0i}$ are used, where $p_i$ is the actual analyte concentration and $p_{0i}$ is the saturation vapor pressure at the measurement temperature [25].

**Table 1.** Properties of analytes, the tested concentration range, saturation vapor pressure at −10 $^0$C

| Analyte | Dielectric constant ($\varepsilon$) | Dipole Moment ($\mu$) | Concentration (ppm) Min. | Max. | $p^0$(-10 $^0$C) ppm |
|---|---|---|---|---|---|
| Toluene (TOL) | 2.38 | 0.26 | 120 | 1200 | 4920 |
| Trichloroethylene(TCE) | 3.40 | 0.81 | 550 | 3100 | 15800 |
| Acetone (ACE) | 6.2 | 2.91 | 1440 | 1152 | 28800 |
| Ethanol (ETH) | 24.50 | 1.69 | 460 | 2300 | 9200 |
| Water | 80.10 | 1.85 | | | |

The response of sensors to the various gas environments is defined as $S = \frac{\Delta R}{R_0} = \frac{R_g - R_0}{R_0}$ where $R_g$ is the resistance when exposed to gas environments, $R_0$ is the resistance of the baseline in the dry air or continued humid air. The $S$ value is represented as sensitivity in the literature; however, this should be defined as the deviation from the baseline.

Sensitivity of sensor can be defined as follows [27];

$$\bar{S} = \frac{1}{nR_0}\sum_{i=1}^{n}\frac{(R_{gi} - R_{g(i-1)})}{(C_{gi} - C_{g(i-1)})} \tag{eq. 4}$$

where $C_{gi}$ represents the concentration for the applied gas, $R_{gi}$ is the measured resistance for the applied gas concentration and n is the number of applied concentrations. If i=1 then $R_{g0} = R_0$ and $C_{g0} = 0$. This approach gives the concentration of the detected gas per one unit changes.

## 3. Results and Discussion

### 3.1 Morphological and physical characterizations

It is well known that the sensing layer plays a crucial role in the sensors and its structural and chemical characteristics will directly affect the sensing performance. TEM analysis (Fig. 2-a) reveal that the nanofibers involve a granular crystalline microstructure with tiny nanograins. In addition, nanofibers include only tin, oxygen and palladium atoms, which demonstrates the absence of the impurities (Fig. 2-b), indicating the successful removal of organics with calcination.

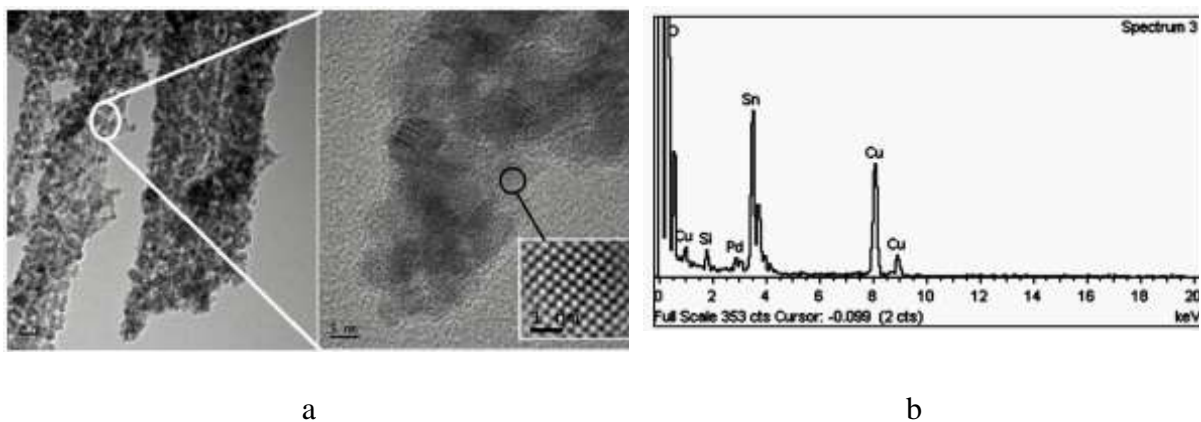

**Figure 2.** a) TEM image and b) EDS spectrum of SnO2-Pd nanofiber

As revealed by SEM analysis (Fig. 3-a), SnO$_2$-Pd nanofiber layer exhibits continuous interconnected network structure with a fiber diameter of 50-100 nm, which proves the successful electrospinning process. In addition, it was demonstrated that MWCNT-COOH layer were homogeneously coated on the nanofiber layer (Fig. 3-b).

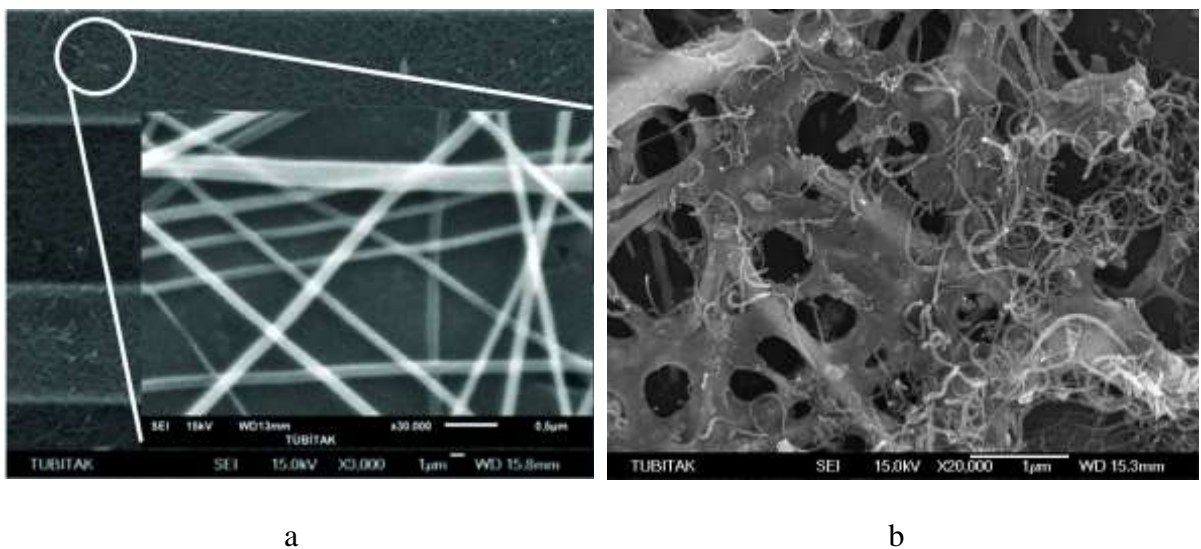

**Figure 3.** SEM image of a) SnO2-Pd nanofiber layer b) MWCNT-COOH layer on nanofibers

On the other hand, crystalline structure of the SnO$_2$-Pd nanofibers was investigated by XRD. It is obvious that nanofibers show polycrystalline structure (Fig. 4-a), demonstrating the successful calcination process (600 ºC) in agreement with previous works[12, 28]. The formation of polycrystalline structure could be attributed to the removal of PVP and oxidation of tin element. No peaks from either any other phase of SnO$_2$ or any impurities were observed. Moreover, TGA analysis was carried out to investigate the thermal degradation behavior of PVP during calcination process. As shown in Fig. 4-b, the weight loss increases dramatically with increasing the temperature and levels off at 400 ºC, which can be explained by the removal of all organics. This again indicates the successful calcination process and formation of pure SnO$_2$ nanofibers[29].

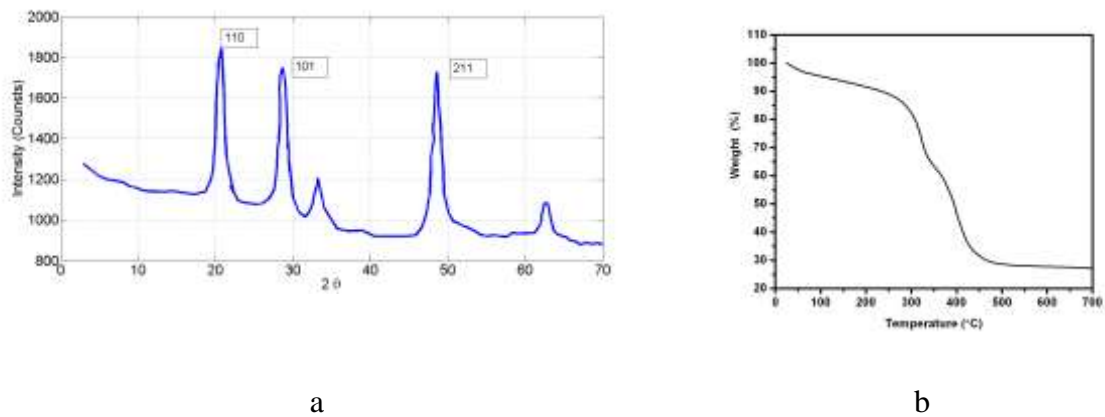

a  b

**Figure 4.** a) XRD spectrum SnO2-Pd nanofiber after calcination and b) TGA analysis of SnO2-Pd nanofiber before calcination at 600 $^0$C.

**3.2 Characterization of sensor responses**

### 3.2.1 Ethanol sensing

Ethanol responses of the SnO$_2$-Pd sensors as a function of temperature were shown in Figure 5a. The sensor response is almost negligible at 200 °C, however, increasing the temperature resulted in better sensor response. On the other hand, the sensor response increased dramatically after coating with MWCNT-COOH layer. This could be explained by the support of carbon nanotubes to the electron transfer from SnO$_2$-Pd nanofibers to the IDE electrodes during sensing (Figure 5b). It can be deduced that all fibers were involved in sensing even if there is no connection between each other or the IDE electrodes.

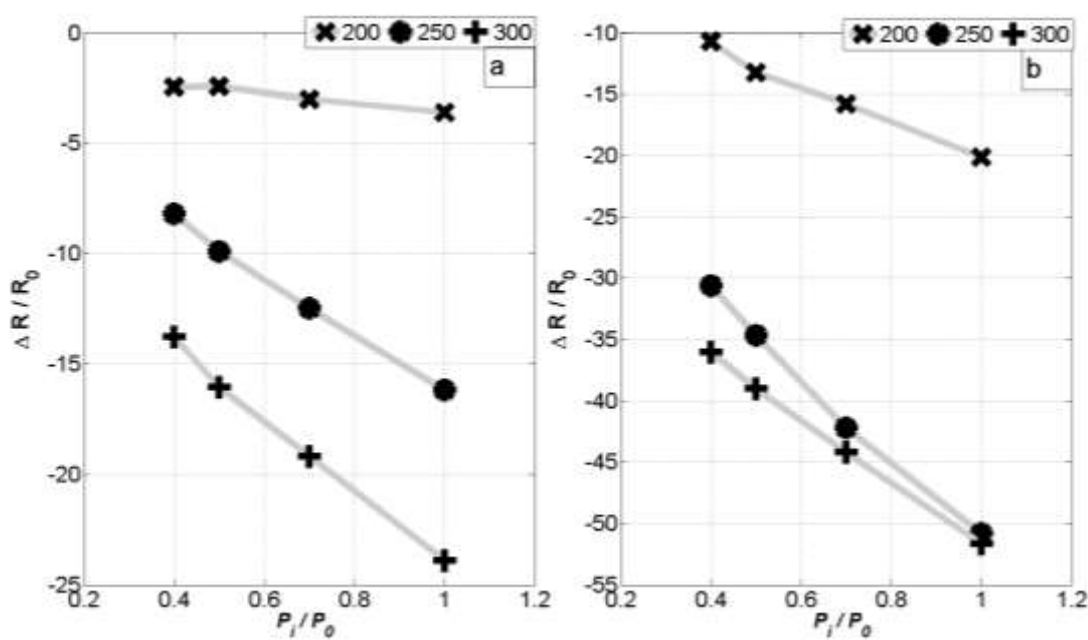

**Figure 5.** Sensor responses of a) SnO$_2$-Pd nanofibers and b) MWCNT-COOH coated SnO$_2$-Pd nanofiber sensors

The sensing mechanism of SnO$_2$-Pd nanofiber system is based on the interaction between the lattice and -OH groups of ethanol. We believe that the detection of ethanol is similar to water. Size of ethanol molecule is larger than the water molecule; therefore, interaction of ethanol occurs on the surface of SnO$_2$-Pd nanofiber and generates the electron according to Eq. 1 and 2.

$$C_2H_5OH + Sn_{lat} + O_{lat} \leftrightarrow (C_2H_5 - Sn_{lat}) + O_{lat}OH + e \qquad (Eq.\ 5)$$

or

$$C_2H_5OH + 2Sn_{lat} + O_{lat} \leftrightarrow 2(C_2H_5OH - Sn_{lat}) + V_0 \qquad (Eq.\ 6)$$

Sensing mechanism of the MWCNT-COOH coated $SnO_2$-Pd nanofiber sensor could be explained by the previous work by Sin *et al*[30]. Owing to its hydroxyl group, ethanol might interact with carboxylic group of carbon nanotubes via hydrogen bonding[24]. This may lead to diminish the number of free charge carriers-holes and thus increase the electrical resistance upon ethanol exposure[30]. It is obvious that the mechanisms mentioned above works opposite to each other. Upon the adsorption of ethanol on the nanofibers, oxide structure generates electrons and injects into the MWCNT-COOH layer. It seems that the number of electrons generated by the nanofibers is higher than the holes that was generated via interaction of ethanol with carbon nanotubes. Hence, an increase in the electrical conductivity was observed for all tests.

Influence of temperature on the electrical conductivity and sensing behavior is another issue that should be considered. Based on the previous works [16, 31], the conductivity of both $SnO_2$-Pd and MWCNT-COOH layers and thus, sensor baseline is expected to increase by increasing the temperature. Nevertheless, sensor responses of the layers are different due to the dissimilar sensing mechanism, as shown in Figure 6.

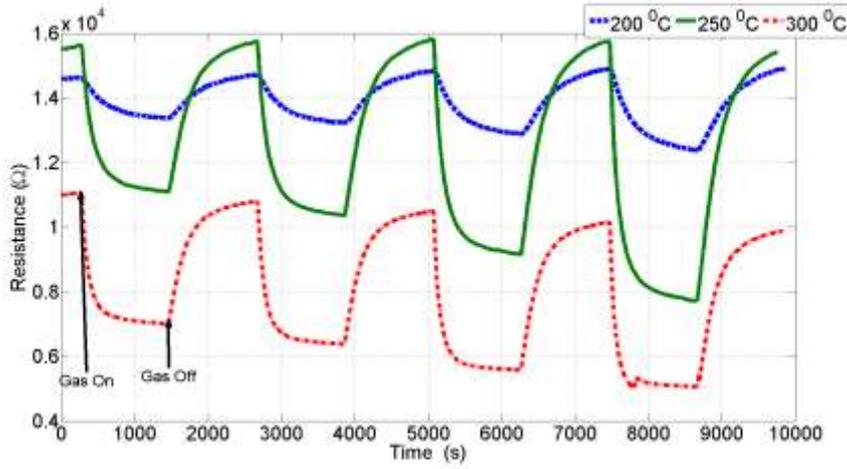

**Figure 6.** Ethanol response of MWCNT-COOH coated $SnO_2$-Pd sensor at 200, 250 and 300 $^0$C.

The change in the conductivity and in the sensing behavior of the composite structure with temperature could be understood with the following explanations. There are two different effects in the sensing mechanism. First one is the positive effect on the electrical conductivity upon the interaction of the ethanol with $SnO_2$ (See Eq. 5 and 6). Second one is the negative effect of the carboxylic group of MWCNT-COOH due to its electronegativity. The carboxylic groups attract the free electrons that interact with $SnO_2$ structure, which in turn led to decrease in conductivity. Therefore, the sensing mechanism of $SnO_2$ nanofiber and composite structure (MWCNT-COOH coated $SnO_2$) is expected to be different.

If there were only $SnO_2$ nanofibers in the system, the conductivity would increase with temperature (Eq. 3) and interaction with the gas (Eq. 5 and 6). However, adding MWCNT-COOH to the system means adding binary structure which both increases and decreases electrical conductivity.

Sum of both cases define the system's total behavior. Hence,

$$\sigma_{CNT-COOH} = \sigma_{T(CNT-COOH)} - \sigma_{S(CNT-COOH)} \quad \text{(Eq. 7)}$$

where $\sigma_{T(CNT-COOH)}$ and $\sigma_{S(CNT-COOH)}$ defines the conductivity change with temperature and interaction with the gases. Minus term demonstrates that the MWCNT-COOH absorbs some of the electrons released after interaction with the gas.

$$\sigma_{SnO_2} = \sigma_{T(SnO_2)} + \sigma_{S(SnO_2)} \tag{Eq. 8}$$

However, Eq. 8 shows that the conductivity increases both with temperature and interaction with the gas. Therefore, the total behavior of the MWCNT-COOH coated SnO$_2$ nanofiber system could be explained as follows;

$$\sigma_{TOTAL} = \sigma_{T(CNT-COOH)} + \sigma_{T(SnO_2)} + \sigma_{S(SnO_2)} - \sigma_{S(CNT-COOH)} \tag{Eq. 9}$$

where $\sigma_{TOTAL}$ is the total conductivity of during gas exposure.

### 3.2.2 Ethanol sensing with humidity

Effect of humidity on the ethanol sensing was also investigated. There is a slight decrease (~1-3 %) in the sensor response when humidity was incorporated in the ethanol, as shown in Figure 7. Since the interaction mechanism of humidity and ethanol with SnO$_2$-Pd structure is similar, most of the interaction sites were occupied by the water molecules, which led to a decrease in ethanol response. As the temperature was increased, it was found that the sensitivity as also increased, which could be explained by the increased number of sensing sites [32].

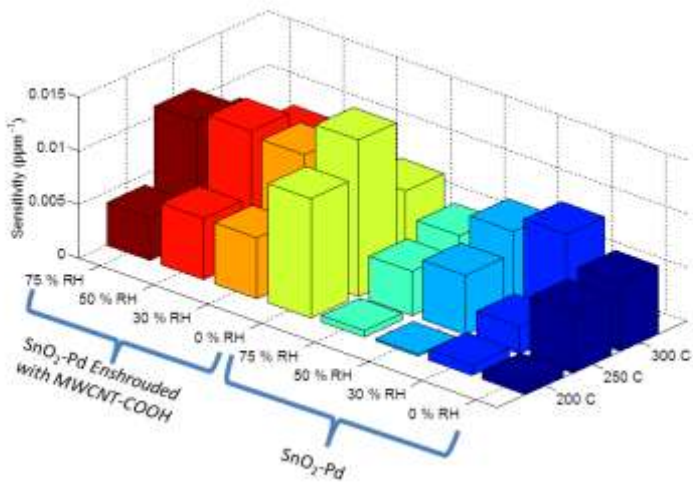

**Figure 7.** Sensitivity values of SnO$_2$-Pd nanofiber layer and MWCNT-COOH coated nanofiber layer against ethanol with different humidity levels (calculated with eq. 4)

The sensing mechanism of the enshrouded SnO$_2$-Pd structures can be explained by eq. 9. Based on this equation, it can be deduced that the generated electrons with temperature increase was absorbed by the MWCNT-COOH layer. In this study, maximum sensitivity against ethanol was achieved at 250 $^0$C for all humidity levels.

nanofiber layer against ethanol with different humidity levels (calculated with eq. 4)

### 3.2.3 Sensing behaviour of other gases

As demonstrated in Figure 8a, when the SnO$_2$-Pd nanofiber sensor was exposed to acetone at 200 ºC, no response was detected. Increasing the temperature has very small influence on the sensor response. This could be attributed to the absence of proper functional group of acetone that could interact with SnO$_2$-Pd surface. The behavior of TOL and TCE was also found to be similar (not shown here).

On the other hand, there are increased responses for MWCNT-COOH coated SnO$_2$-Pd nanofiber layer (Figure 8b and Figure 9b). Carbon nanotubes helps the electron transfer generated during sensing process and sensitivity values increase order of ~10$^3$ (Compared to Fig. 8-a and 9-a). Considering these issues, it can be concluded that MWCNT layer works as a filter for the underlying SnO$_2$ nanofiber layer as well as it works as a sensing layer.

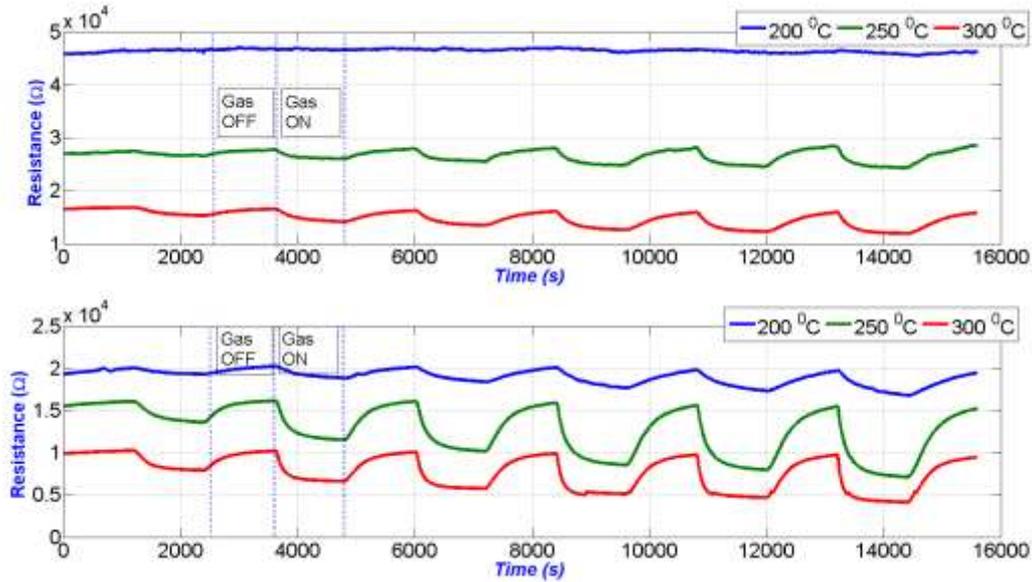

**Figure 8.** Acetone responses for two sensor structures at different temperature a) $SnO_2$-Pd nanofiber and b) Enshrouded $SnO_2$-Pd nanofiber with MWCNT-COOH.

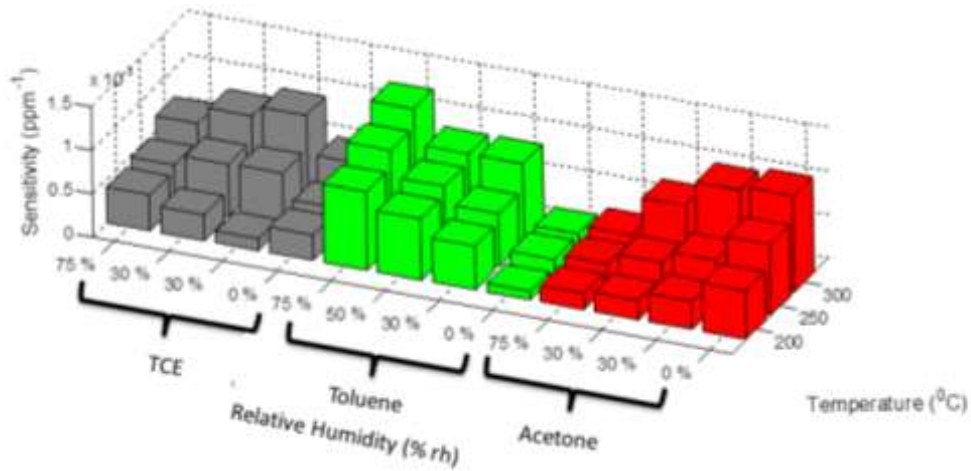

**Figure 9.** Sensitivities of the sensors at different temperature and humidity levels **a)** $SnO_2$-Pd nanofiber **b)** Enshrouded $SnO_2$-Pd nanofiber with MWCNT-COOH.

## Conclusions

In this work, preparation of ethanol sensor based on $SnO_2$-Pd/MWCNT-COOH composite structure was successfully achieved. Increasing the temperature and gas concentration has led to increased sensor response for $SnO_2$-Pd nanofiber system. However, there is no general rule for MWCNT-COOH coated $SnO_2$-Pd nanofiber system. Sensing mechanisms was modeled for electrical charge transfer between nanofibers and nanotubes. We have systematically investigated the effect of sensing layer composition on the sensing mechanism.

Acknowledgement

The author would like to acknowledge the financial support from TUBİTAK (Grant No:115E045).